\documentclass[onecolumn]{emulateapj}
\usepackage{lscape}

\shorttitle{Suzaku Observation of MeV blazar J0746}
\shortauthors{Watanabe et al.}

\begin{document}

\title{Suzaku Observations of extreme MeV blazar SWIFT J0746.3+2548}

\author{Shin Watanabe\altaffilmark{1,2},
Rie Sato\altaffilmark{1},
Tadayuki Takahashi\altaffilmark{1,2},
Jun Kataoka\altaffilmark{3},
Greg Madejski\altaffilmark{4},\\
Marek Sikora\altaffilmark{5},
Fabrizio Tavecchio\altaffilmark{6},
Rita Sambruna\altaffilmark{7},
Roger Romani\altaffilmark{4},
Philip G. Edwards\altaffilmark{8,1} \\ and
Tapio Pursimo\altaffilmark{9}}
\email{watanabe@astro.isas.jaxa.jp}

\altaffiltext{1}{Institute of Space and Astronautical Science/JAXA, Sagamihara, Kanagawa 229-8510, Japan}
\altaffiltext{2}{Department of Physics, University of Tokyo, Bunkyo, Tokyo, 113-0033, Japan}
\altaffiltext{3}{Department of Physics, Tokyo Institute of Technology, Meguro, Tokyo, 152-8551, Japan}
\altaffiltext{4}{Kavli Institute for Particle Astrophysics and Cosmology, Stanford University, Stanford, CA 94305, USA}
\altaffiltext{5}{Nicolaus Copernicus Astronomical Center, Bartycka 18, 00-716, Warsaw, Poland}
\altaffiltext{6}{INAF-Osservatorio Astronomico di Brera, Via Bianchi 46, l-23807, Merate (LC), Italy}
\altaffiltext{7}{NASA Goddard Space Flight Center, Greenbelt, MD 20771, USA}
\altaffiltext{8}{Australia Telescope National Facility, CSIRO, Locked Bag 194, Narrabri NSW 2390, Australia}
\altaffiltext{9}{Nordic Optical Telescope, Apdo 474, 38700 Santa Cruz de La Palma, Spain}

\begin{abstract}
We report the $Suzaku$ observations of the high luminosity blazar SWIFT J0746.3+2548 (J0746) 
conducted in November 2005.  This object, with $z = 2.979$, is 
the highest redshift source observed in the 
$Suzaku$ Guaranteed Time Observer (GTO) period, is likely to show 
high gamma-ray flux peaking in the MeV range.
As a result of the good photon statistics and high signal-to-noise ratio spectrum, 
the $Suzaku$ observation clearly confirms that J0746 has an extremely hard spectrum in the energy
range of 0.3--24~keV, 
which is well represented by a single power-law with a photon index of 
$\Gamma_{\rm ph} \simeq 1.17$ and Galactic absorption.  The multiwavelength 
spectral energy distribution of 
J0746 shows two continuum components, and is well modeled assuming that the high-energy 
spectral component results from Comptonization of the broad-line region photons.  
In this paper we search for the bulk Compton spectral features predicted to be 
produced in the soft X-ray band by scattering external optical/UV photons by cold 
electrons in a relativistic jet. We discuss and provide constraints on the pair 
content resulting from the apparent absence of such features.
\end{abstract}


\section{Introduction}
Blazars are a sub-category of Active Galactic Nuclei (AGN) whose jet 
emission is pointing close to our line of sight \citep[e.g.,][]{urr95, ulr97}. 
Generally, the overall 
spectra of blazar sources (plotted in the log$(\nu)$-log$(\nu F_{\nu})$ 
plane, where $F_{\nu}$ is the observed spectral flux energy density)
have two pronounced continuum components: one peaking between IR and
X-rays and the other in the $\gamma$-ray regime \citep[see, e.g.,][]{kub98, ghi98}. 
The lower energy component is believed
to be produced by the synchrotron radiation of relativistic electrons 
accelerated within the outflow, while inverse Compton (IC) emission by 
the same electrons is most likely responsible for the formation of the 
high energy $\gamma$-ray component.  The spectral energy distributions 
(SEDs) of blazars form a sequence in luminosity, with more luminous 
sources having both peaks at lower energies than their fainter 
counterparts \citep{fos98,ghi98}.  In this 
sequence, flat-spectrum radio quasars (FSRQs) are the most luminous objects.
It is widely believed, in addition, that the IC emission from FSRQs 
is dominated by the scattering of soft photons external to the jet 
(external Compton process, ERC).   Those photons, in turn, are produced 
by the accretion disk, and interact with the jet either directly or 
indirectly, after being scattered or reprocessed in the broad-line region 
(BLR) or a dusty torus 
\citep[see, e.g.,][]{der93,sik94}. 
Other sources of seed photons can also contribute to the 
observed IC component;  in particular, those can be the synchrotron photons 
themselves, radiating via the synchrotron self-Compton (SSC) process 
\citep{sok05}. 
In FSRQs, the synchrotron emission peaks around IR frequencies, and thus 
the nonthermal X-ray emission is relatively weak compared to that of other 
types of blazar sources.  These spectral features suggest that FSRQs are 
well-suited for searching for the bulk Compton (BC) spectral component, 
which is produced by Comptonization of external UV radiation by cold 
electrons in a jet 
\citep{beg87,sik00,mod04,cel07}.
Using $Suzaku$ data of PKS1510$-$089, \citet{kat08} argued that the observed soft X-ray 
excess below 1~keV and the plausible one at $\sim 18$~keV may be such bulk-Compton 
features produced by inhomogeneities prior to their collision and shock formation 
(the latter being the site of particle acceleration and production 
of the nonthermal radiation).

SWIFT J0746.3+2548 (J0746) was identified with an optically faint 
quasar at $z = 2.979$ detected in the 15--200~keV energy band the 
Burst Alert Telescope \citep[BAT;][]{bar05} on board the $Swift$ 
satellite.  Its broadband spectrum is representative of other FSRQs, 
which have two continuum components: one peaking at IR wavelengths 
and the other at MeV energies.  This qualifies J0746 as a likely new member 
of the MeV blazar class \citep{sam06}.  The X-ray spectrum observed 
by $Swift$ XRT was hard in the 0.5--8 keV with a photon index $\Gamma_{\rm ph} 
\sim 1.3$. Moreover, $Swift$ data showed spectral hardening at energies $<$1~keV, 
which \citet{sam06} interpreted as an excess absorption over the 
Galactic value, or a flatter power-law component, implying a sharp 
($\Delta\Gamma_{\rm ph}\sim1.1$) spectral break at $\sim$4~keV.  It has been 
argued that a clear distinction between the two possibilities can be made 
by $Suzaku$, due to its broad bandpass, good photon statistics and low background data.

In 2005 November, simultaneous observations were performed with the Very Large Array 
(VLA; radio), the 26 m diameter University of Michigan Radio Astronomy Observatory 
(UMRAO; radio), the 14 m diameter Mets$\ddot{a}$hovi radio-telescopes (radio), 
the Hobby-Eberly Telescope (HET; optical), the 2.56m Nordic Optical Telescope(NOT; 
optical), $Swift$ XRT and UVOT (optical-UV, X-ray) and $Suzaku$ (X-ray). 
\citet{sam06} presented some of the simultaneous observations in addition 
to the description of J0746 discovery with $Swift$. In this paper, we report 
a detailed analysis of the $\sim 100$~ks $Suzaku$ observation of J0746 as a part 
of the SWG (science working group) program. Moreover, we present the multiband 
analysis using all simultaneous observations. This paper is organized as follows.
In $\S$2, we described the $Suzaku$ observation and data reduction.  In $\S$3, 
we report the optical and radio results, which were not reported 
in \citet{sam06}. In $\S$4, we present the spectral analysis of the $Suzaku$ 
X-ray data as well as multiband analysis.  Finally, we discuss the constraints on the 
content of the jet inferred from the X-ray spectrum in $\S$5.  Throughout this 
paper, we adopt  the cosmological parameters $H_0 = 71$ km s$^{-1}$ Mpc$^{-1}$, 
$\Omega_{\rm M} = 0.27$ and $\Omega_{\Lambda} = 0.73$.

\section{Suzaku Observation and Data Reduction}

\subsection{Observation}

J0746 was observed with $Suzaku$ \citep{mit07} from 2005 November 4 at 08:20 UT 
until November 6 14:04 UT, during the performance verification (PV) phase. 
Table 1 summarizes the start and end times, and the exposures of the $Suzaku$ observation 
(sequence number 700011010). $Suzaku$ has four sets of X-ray telescopes \citep{ser07}
each with a focal-plane X-ray CCD camera \citep[X-ray Imaging Spectrometer(XIS);][]{koy07} 
that are sensitive in the energy range of 0.3--12~keV. 
Three of the XIS (XIS 0, 2, 3) detectors have front-illuminated (FI) CCDs, 
while the XIS 1 utilizes a back-illuminated (BI) CCD. 
The merit of the BI CCD is its improved sensitivity in the soft X-ray energy band below 1~keV.
$Suzaku$ also features a non-imaging collimated Hard X-ray Detector \citep[HXD;][]{tak07}, 
which covers the 10--600~keV energy band with Si PIN photodiodes and GSO scintillation detectors. 
$Suzaku$ has the two default pointing positions, XIS nominal position and HXD nominal position.
In this observation, we used the HXD nominal position, in which the effective area of the HXD
is maximized, whereas that of the XIS is reduced to $\sim$~88\% on the average.

\subsubsection{XIS Data Reduction}

The XIS data used in this paper were version 1.2 of the cleaned data.
The screening is based on the following criteria: 
(1) ASCA-grade 0, 2, 3, 4, and 6 events were accumulated, and the CLEANSIS script was used to remove 
hot or flickering pixels, 
(2) data collected within 256~s of passage through the South Atlantic Anomaly (SAA) 
were discarded, and 
(3) data were selected to be 5\arcdeg ~ in elevation above the Earth rim
(20\arcdeg ~ above the day-Earth rim).
After this screening, the net exposure for good time intervals is 100.5~ksec.

The XIS events were extracted from a circular region with a radius of 2.6\arcmin ~
centered on the source peak, whereas the background was accumulated in an annulus with inner 
and outer radii of 2.6\arcmin ~ and 4.3\arcmin, respectively.
The response (RMF) files used in this paper are the standard RMF files 
(ae\_xi\{0,1,2,3\}\_20060213.rmf), provided by the XIS instrumental team.
The auxiliary (ARF) files are produced using the analysis tool XISSIMARFGEN developed 
by the $Suzaku$ team, which is included in the software package HEAsoft version 6.2.

\subsubsection{HXD/PIN Data Reduction}

The PIN source spectra were extracted from cleaned version 1.2 HXD/PIN event files.
Data were selected according to the following criteria:
(1) More than 500~s from a South Atlantic Anomaly passage, (2) cut-off rigidity above 8~GV, and 
(3) day- and night-Earth elevation angles each 5\arcdeg. 
After this screening and the dead time correction using "pseudo-events" \citep{kok07}, 
the net exposure for good time intervals becomes 74.0~ksec.

The PIN spectrum is dominated by the time-variable instrumental background 
induced by cosmic-rays and trapped charged particles in the satellite orbit.
The HXD instrument team has developed an effective method \citep{wat07}
of modeling the time-dependent non-X-ray background (NXB) by making use of the PIN 
upper discriminator (UD) signal that monitors passing charged particles 
through the silicon PIN diode. The background spectrum is generated based on a database
of NXB observations accumulated to date during night- and day-earth observations.
The current NXB model is shown to be accurate within $\sim$~4\%. \citep{miz06}.

Another component of the HXD PIN background is the Cosmic X-ray background (CXB).
The form of the CXB was taken as 
9.0$\times$10$^{-9}$(E/3~keV)$^{-0.29}$$\exp$($-$E/40~keV)~erg~cm$^{-2}$~s$^{-1}$ \citep{gru99}.
The CXB spectrum observed with HXD/PIN was simulated by using a PIN response file 
for isotropic diffuse emission (ae\_hxd\_pinflat\_20060809.rsp). 
However, \citet{kok07} reported that the PIN returns a $\sim$13--15\% larger 
normalization than the XIS based on the most recent calibration using the Crab Nebula.
Additionally, it was reported that the XIS normalization of the Crab Nebula agrees 
with the conventional Crab Nebula flux derived from previous satellites. 
Therefore, we introduced a scaling factor of 1.13 to normalize the CXB spectrum.

We used the response files version ae\_hxd\_pinxinom\_20060814.rsp, provided by
the HXD instrumental team. As reported in \citet{kok07}, the response file 
returns 15\% larger flux at the HXD nominal position pointing than the XIS flux. 
Therefore, we corrected the normalization of the HXD/PIN in the spectral analysis ($\S$3.2).

\begin{figure}
  \begin{center}
\epsscale{.80}
\plotone{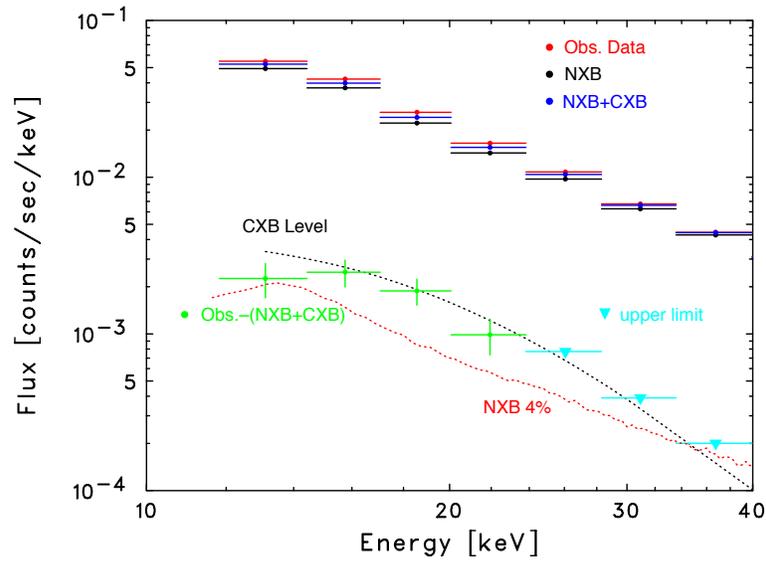}    
  \end{center}
  \caption{The time averaged HXD/PIN spectra. The red and the black show the observed data and  
  the non X-ray background (NXB) model spectrum, respectively. The background model spectrum 
  including NXB and CXB is plotted in blue. After the background subtraction, the detected spectrum
  and the upper limit assuming the 4\% accuracy of the NXB model are plotted in green and cyan, respectively.
  \label{fig:pin_spec}}
\end{figure}

Figure~\ref{fig:pin_spec} shows the time averaged HXD/PIN spectrum.  The NXB model, 
the CXB model and the 4\% level of the NXB are also plotted in the spectrum. 
The hard X-ray emission of J0746 was detected in the energy range from 12~keV
to 24~keV, assuming the 4\% accuracy of the current NXB model.  Above 24~keV, the 
upper limit of flux could be derived from the accuracy of the NXB model.  
We also note here that the source was not detected in the GSO data.  

\section{Optical and Radio Observations}

\subsection{Hobby-Eberly Telescope (HET)}

\begin{figure}
\begin{center}
\epsscale{.70}
\plotone{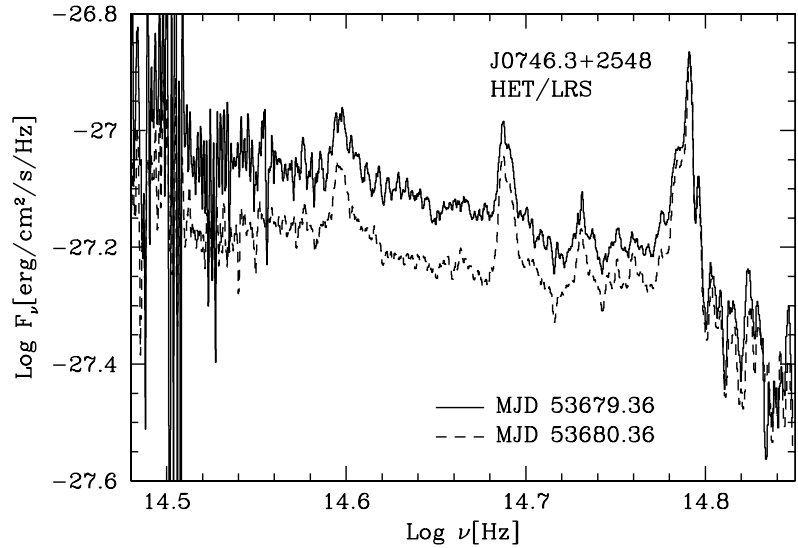}
\caption{Optical spectra of J0746 during 2005 November observations.
\label{fig:opt_spec}}
\end{center}
\end{figure}

The optical spectra were obtained with the 9.2m Hobby-Eberly telescope \citep[HET;][]{ram98}
Marcario Low Resolution Spectrograph \citep[LRS;][]{hil98}.
Observations were made from 2005 November 5 to November 6 
covering $\lambda$420--1000~nm at 1.6~nm resolution. The seeing was
variable during the integrations and so spectrophotometry was not
attempted. However, observations were taken with the slit along the
parallactic angle and at constant air mass and, using white-light pre- and post-
spectrum direct images, indicating that the source faded by 0.17 magnitudes
between the observations.  Using this information, we correct for the
differential slit losses, adjusting the first (worse-seeing) spectrum to
that of November 6. The resulting spectra are shown in Figure \ref{fig:opt_spec}, 
after standard calibrations and correction for an estimated Galactic extinction
of E(B$-$V)=0.07.
 
Overall, the spectra do not differ dramatically from the 
spectrum of the blazar available from the SDSS archive. However, there
is clearly a fading component in the continuum. In the first spectrum,
a power-law fit to the continuum to the red of Ly$\alpha$ gives
$F_\nu=6.6\times 10^{-28}(\nu/10^{14.7}{\rm Hz})^{-0.8}{\rm erg/cm^2/s/Hz}$;
during the second observation the continuum flux was
$F_\nu=5.6\times 10^{-28}(\nu/10^{14.7}{\rm Hz})^{-0.6}{\rm erg/cm^2/s/Hz}$,
the residual slit losses leave a $\sim 10$\% uncertainty in the absolute fluxes.
The broad-line flux is, as expected, nearly constant on this timescale,
confirming the relative spectral normalization estimated from the direct images.
The difference spectrum between the two epochs is nearly pure continuum,
with a spectral index of $\alpha$~$\sim$~2--2.5, suggesting that a fading
synchrotron component is contributing to the optical flux during the
tail of the outburst.  These two observations were obtained only a month after the 
observations reported by \citet{sam06}, conducted on 2005 October 10 - also 
with the LRS on the HET - reported by \citet{sam06}, and the continuum flux
(at 6500 angstrom) is about twice of the average of the two of our flux measurements 
taken in November 2005.

\subsection{Nordic Optical Telescope (NOT)}

We carried out the photometric observations of J0746 at the 2.56 m Nordic Optical Telescope (NOT) 
at La Palma (Observatorio del Roque de los Muchachos, Canary Islands) 
on 2005 November 5 using ALFOSC (Andalucia Faint Object Spectrograph and Camera).
The data have been reduced using the "standard" IRAF procedures
(de-biasing and flat-fielding were applied for all images) 
and the magnitudes were measured using IRAF/apphot-package. 
The magnitudes of the object and the comparison stars 
were measured using a relatively small aperture (about the size of the seeing disc)
and the final brightness of the object via differential photometry.
In order to flux calibrate the comparison stars, the magnitudes were measured 
using a large aperture (19\arcsec diameter).
The brightnesses of the comparison stars were determined using
two different techniques: 
(1) using the published SDSS g'and r' magnitudes and transforming these magnitudes to V and R, respectively, 
(2) using Landolt standard stars (PG2213$-$006 and Mark\_A) \citep{lan92} observed earlier the night having 
about the same airmass as the target. 
Galactic extinction was corrected by using \citet{sch98}. 
The R-band and the V-band photometric measurements were made 
with an exposure of 200~seconds. The R-band and V-band fluxes of J0746 were 
18.9 and 19.2 magnitudes, respectively. The detailed results are given in Table~\ref{tab:notobs}.

\subsection{The Very Large Array (VLA)}

We observed J0746 with the Very Large Array (VLA) between 11:25 and
13:25 UT on 2005 November 3, at the end of the reconfiguration period from
DnC array to D array. In D configuration, the most compact array, the
maximum baseline is 1.03~km. A total of 22 antennas were available for the
observations, which were made at 1.425, 4.860, 8.460, 14.940, 22.460 and
43.340~GHz. 3C138 (J0521+1638) was used as the primary flux density
calibrator. J0746 was observed in three blocks, with each block
containing a $\sim$170 second scan at each frequency. Between each block,
a similar block of observations of 3C138 was carried out, using scans of
100 seconds. The 15, 22 and 43~GHz scans, for both J0746 and 3C138,
were preceded by a 270 second pointing scan on source at 8~GHz to
determine a pointing offset for the higher frequencies, following the
standard VLA reference pointing procedure.

The data were amplitude calibrated in AIPS using the scans on 3C138 to set
the flux density scale. At the three highest frequencies the source models
for 3C138, supplied with the data, were used. After amplitude calibration
the data were written out and read into {\tt Difmap}. After initial phase
self-calibration, the data were modeled by a point source. Inspection of
the correlated flux density as a function of ($u,v$) distance confirmed
this assumption was valid at all frequencies for J0746 in this array
configuration. The individual scans were also independently model-fit, but
no evidence for significant variability over the $\sim$2 hour period was
apparent. The results are given in Table~\ref{tab:vlaobs}.

The 1.4~GHz flux density is almost 10\% less than measurements from
$\sim$10 years earlier in the NVSS \citep{con98} and FIRST \citep{bec95} catalogs, however the 4.9~GHz
flux density is significantly higher than the 0.48$\pm$0.04~Jy in the GB6 \citep{gre96}
catalog. The inverted spectrum between 1.4 and 4.9~GHz, $\alpha =$0.56
($S \propto \nu^{+\alpha}$ suggests the presence of strong self-absorption
at the lowest frequency. Above 4.9~GHz the spectral index is $-$0.40,
although the 15~GHz and, to a lesser extent, 22~GHz flux densities
deviate from a single spectral index fit.

\section{Analysis and Results}

\subsection{Temporal analysis}

\begin{figure}
\begin{center}
\epsscale{.80}
\plotone{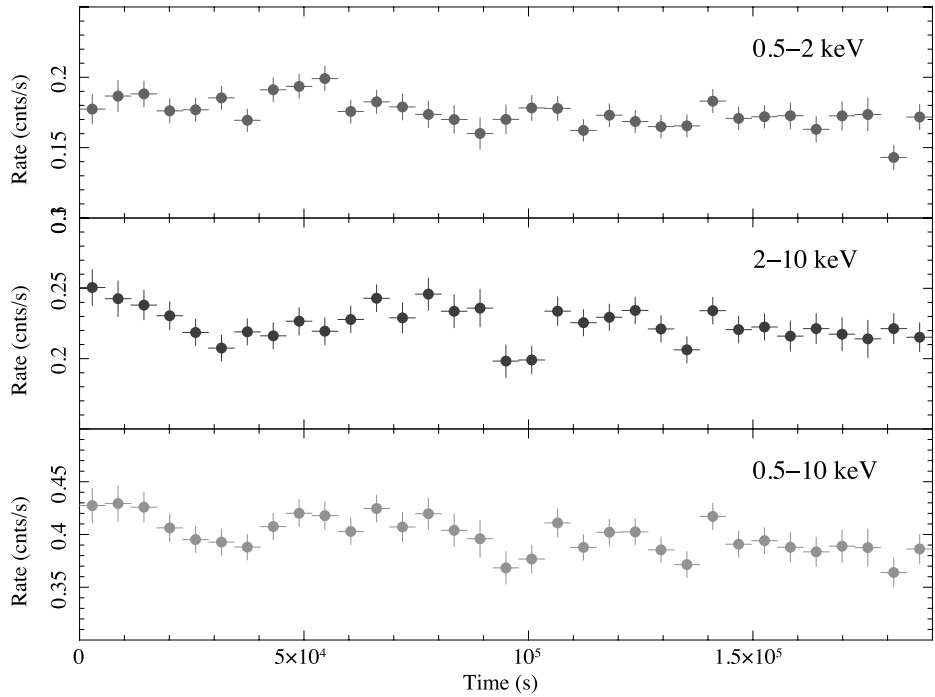}
\caption{Light curves of J0746 during 2005 November observations in the three energy bands: 
0.5--2~keV ($upper$), 2--10~keV ($middle$), and total 0.5--10~keV ($bottom$). 
All the light curves were binned at 5760 s, corresponding to the period of the Suzaku orbit.
\label{fig:xis_lc}}
\end{center}
\end{figure}

\begin{figure}
\begin{center}
\epsscale{.80}
\plotone{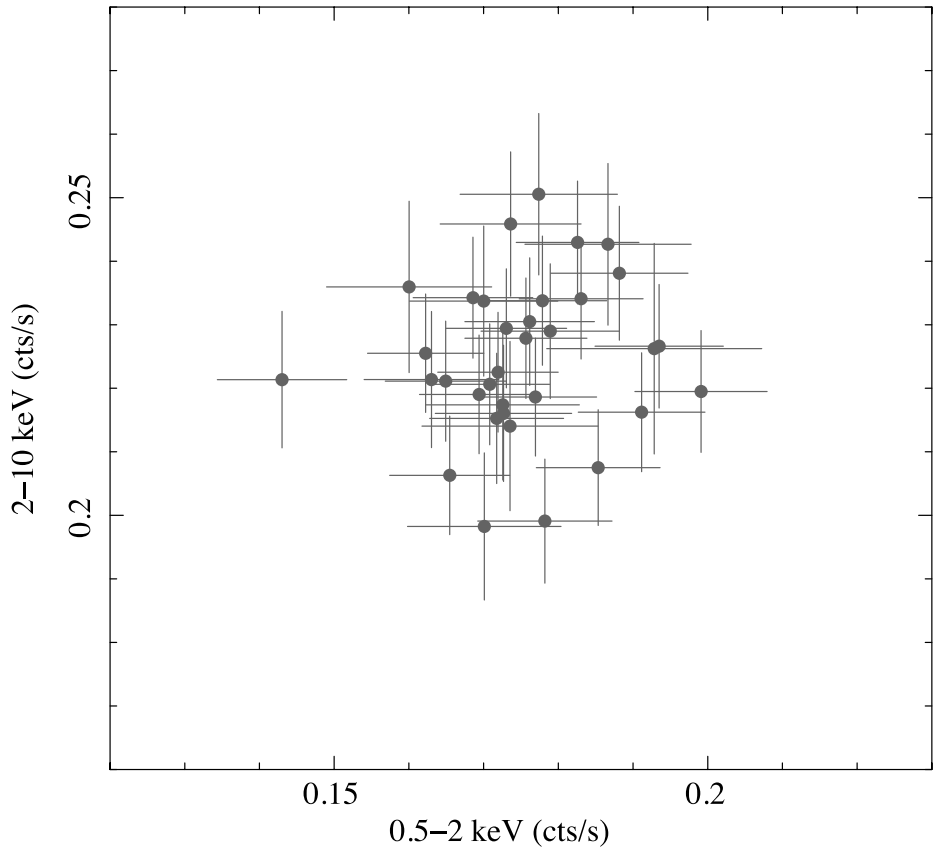}
\caption{Correlation of XIS/FI count rates between 0.5--2~keV and 2--10~keV. 
\label{fig:corr}}
\end{center}
\end{figure}

Figure~\ref{fig:xis_lc} shows the averaged light curves of the XIS/FIs 
in three energy bands: 0.5--2~keV ($upper$ panel), 2--10~keV ($middle$ panel), 
and total (0.5--10~keV; $bottom$ panel), respectively. 
Since the count rate variations of the HXD/PIN detector were less clear due to limited 
photon statistics and uncertainly of the NXB modeling, in the following we concentrate on the temporal variability 
of the XIS data only, below 10~keV.  
Figure~\ref{fig:corr} compares the count-rate correlation between 
the soft X-ray (0.5--2~keV) and the hard X-ray energy bands (2--10~keV). 
We can see that there is no significant correlation between the soft X-ray flux and 
the hard X-ray flux, which indicates that the variability in the soft and hard X-ray bands 
are not well synchronized.
Although $Suzaku$ X-ray light curve shows some variability, it is not nearly 
as strong as that measured by $Swift$, where \citet{sam06} reported 
that the $Swift$ XRT detected a factor of $\sim2$ flux change in a few hours.  
We try to evaluate the variability by calculating the variability amplitude relative to the mean 
count rate corrected for effects of random errors \cite[e.g.,][]{ede02}: 
$F_{\rm var} = 1/\overline{x} \sqrt{S^2 - \overline{\sigma_{\rm err}}^2}$,
where $S^2$ is the total variance of the light curve, 
$\overline{\sigma_{err}}^2$ is the mean error squared and $\overline{x}$ 
is mean count rate.
The variability amplitude of J0746 is $F_{\rm var,soft} \sim 0.033\pm0.018$ 
and $F_{\rm var,hard} \sim 0.026\pm0.013$, and the energy-dependence of variability is flat. 

\subsection{Spectral Analysis}

The XIS and HXD/PIN background subtracted spectra were fitted using XSPEC v11.3.2,
including data within the energy range 0.3--24~keV. The Galactic absorption toward
J0746 is taken to be $N_H$~=~4.04~$\times$~10$^{20}$~cm$^{-2}$ \citep{dic90}.
Note that our best-fitting value for the column density, $N_H=(4.89\pm0.50)\times10^{20}$~cm$^{-2}$ (Table 4), 
which is approximately consistent with the Galactic value and there is no significant excess absorption. 
All errors are quoted at the 68.3\% (1$\sigma$) confidence level for the parameters.
The fits are restricted to the energy range 0.5--10~keV (XIS 0, 2, 3: FI chips), 
0.3--7~keV (XIS 1: BI chip) and 11--24~keV (HXD/PIN), respectively. 
In the following analysis, we fixed the relative normalization of the XISs and PIN at 1.15 
(see $\S$ 2.1.1).

\begin{figure}
\begin{center}
\epsscale{1.}
\plottwo{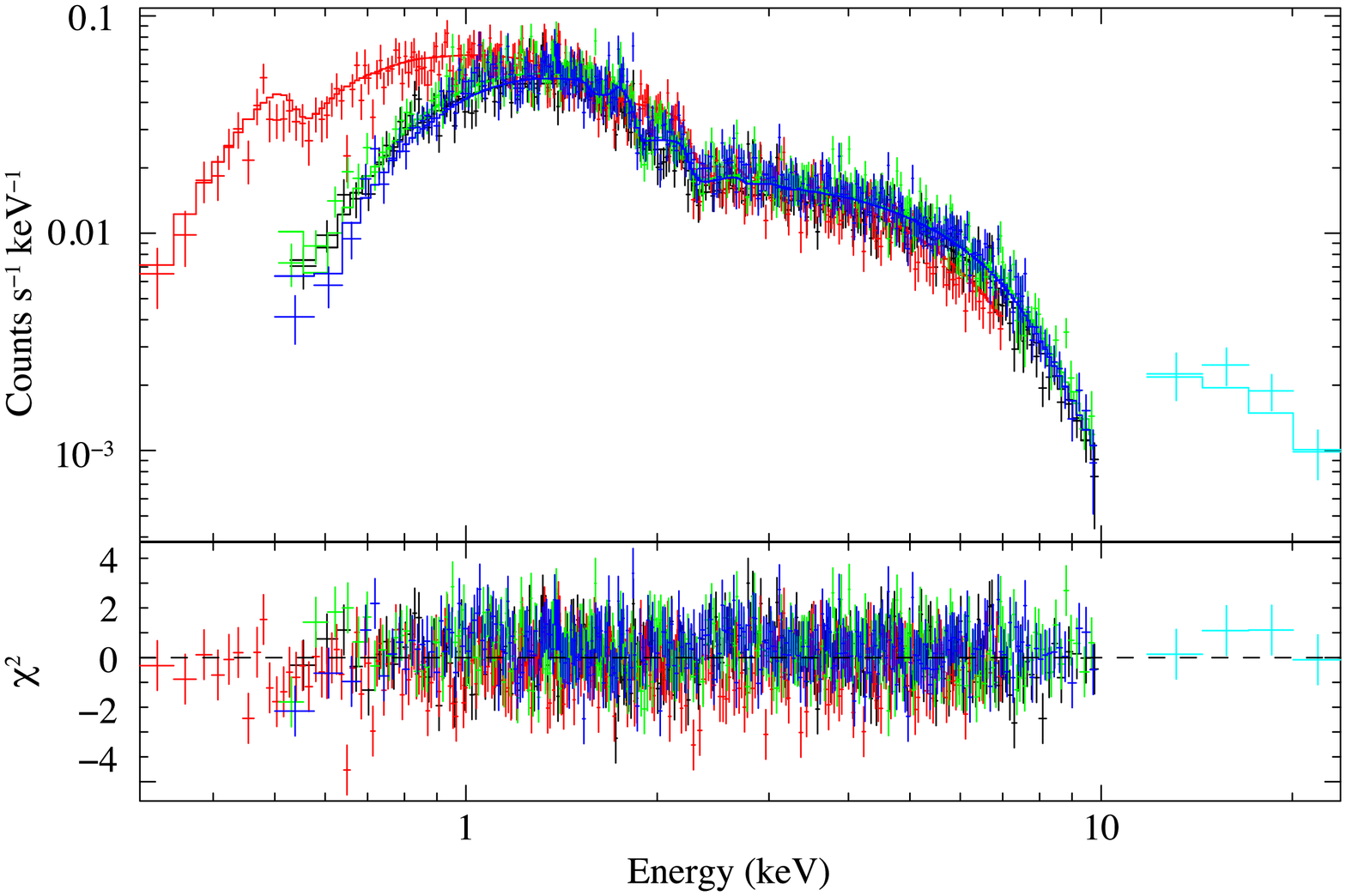}{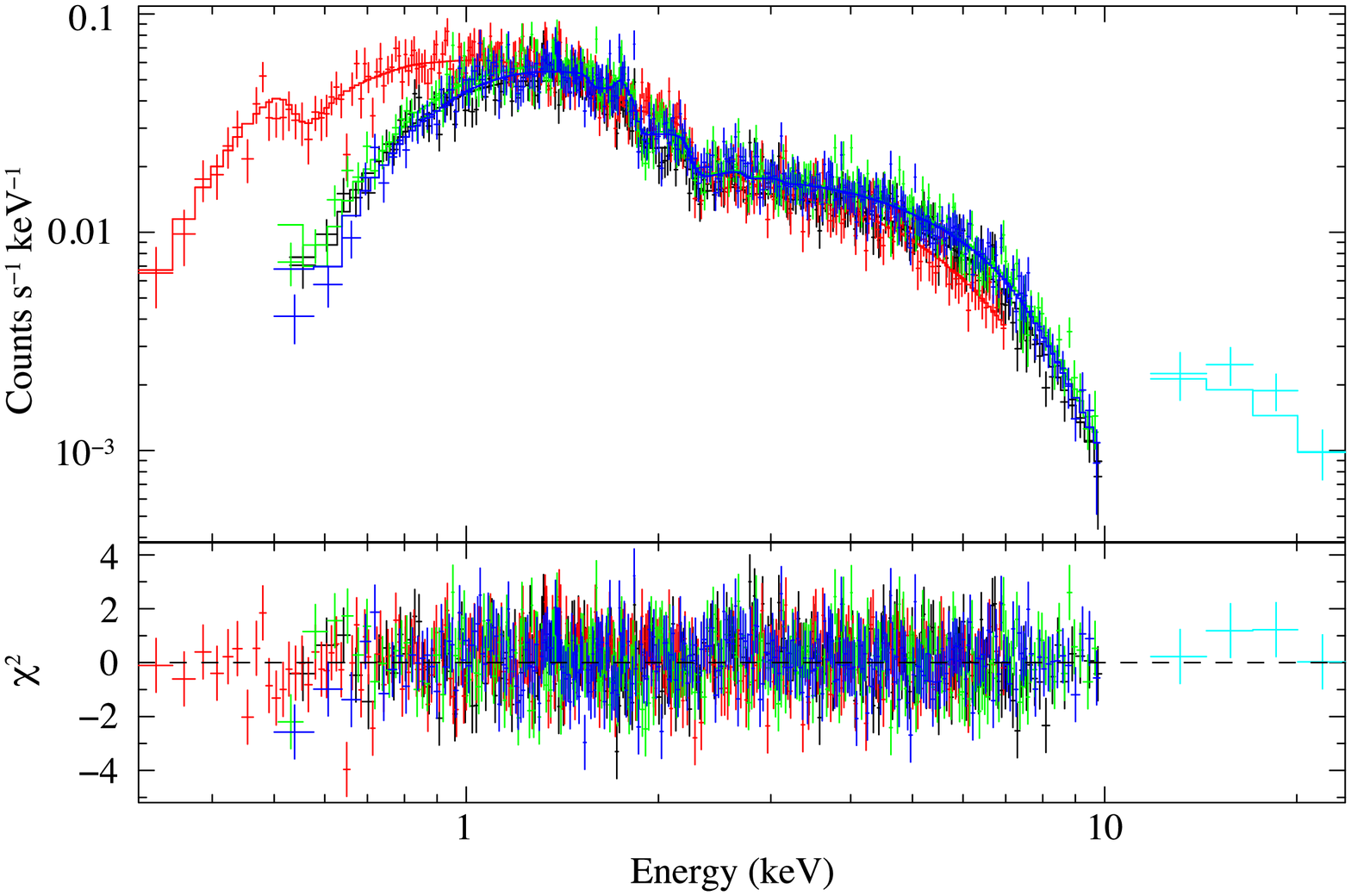}
\caption{$Left$: Broadband (0.3--24~keV; XIS[0--3] + HXD/PIN) Suzaku spectra of J0746. 
The black, red, green and blue points show the XIS0,1,2,3 spectra, respectively. 
The cyan points are HXD-PIN spectrum.
The upper panel shows the background subtracted spectra, plotted with an absorbed power-law model 
of photon index $\Gamma_{\rm ph}=1.17$ and a column density 4.04~$\times$~10$^{20}$~cm$^{-2}$ 
(Galactic value). 
The lower panel shows the residuals to this power-law model fit. Some scatter in the residual
panel shows that spectral normalization is not consistent. 
$Right$: the spectrum plotted against the best-fit model composed of an absorbed power-law 
with constant factors.
  \label{fig:xispin_fit_spec}}
\end{center}
\end{figure}

Figure~\ref{fig:xispin_fit_spec} ($left$) shows four XISs and HXD/PIN background-subtracted 
spectra with residuals to the power-law with the Galactic column density, determined using the data 
from 0.3~keV to 24~keV (model 1). We obtained the photon index of $\Gamma_{\rm ph}$=1.17, 
but this model did not represent the spectra well yielding a $\chi^{2}$/dof of 1238/1112.
Some scatter in the residual panel in Figure~\ref{fig:xispin_fit_spec} ($left$) indicates 
that the spectral normalization among the XISs is not constant. 
To represent the shape of the observed X-ray spectra, we adjusted the normalization factor 
among the four XISs relative to XIS 0 (model 2).
Since \citet{ser07} reported that the spectral normalizations are slightly (a few percent) 
different among the CCD sensors based on the contemporaneous fit of the Crab spectra, 
the few percent adjustment of the relative normalization is reasonable. 
This model well reproduced the spectra with the best $\chi^{2}$/dof of 1113/1109 
(Figure~\ref{fig:xispin_fit_spec}: $right$).
For this model the photon index is $\Gamma_{\rm ph} = 1.18$ with Galactic absorption, and 
the 2--10~keV flux of XIS0 is (3.07~$\pm$~0.03)~$\times$~10$^{-12}$~erg~cm$^{-2}$~s$^{-1}$. 
This corresponds to the $Swift$ XRT flux of $\sim3\times10^{-12}$~erg~cm$^{-2}$~s$^{-1}$. 
All of the spectral fitting results are summarized in Table~\ref{tab:spec}.
We conclude that the X-ray spectra of J0746 within the energy range 0.3--24 keV 
are well described by an extremely hard power-law ($\Gamma_{\rm ph}=1.17$) with the Galactic absorption. 

\subsection{Spectral energy distribution}
 
\begin{figure}
\begin{center}
\epsscale{0.80}
\plotone{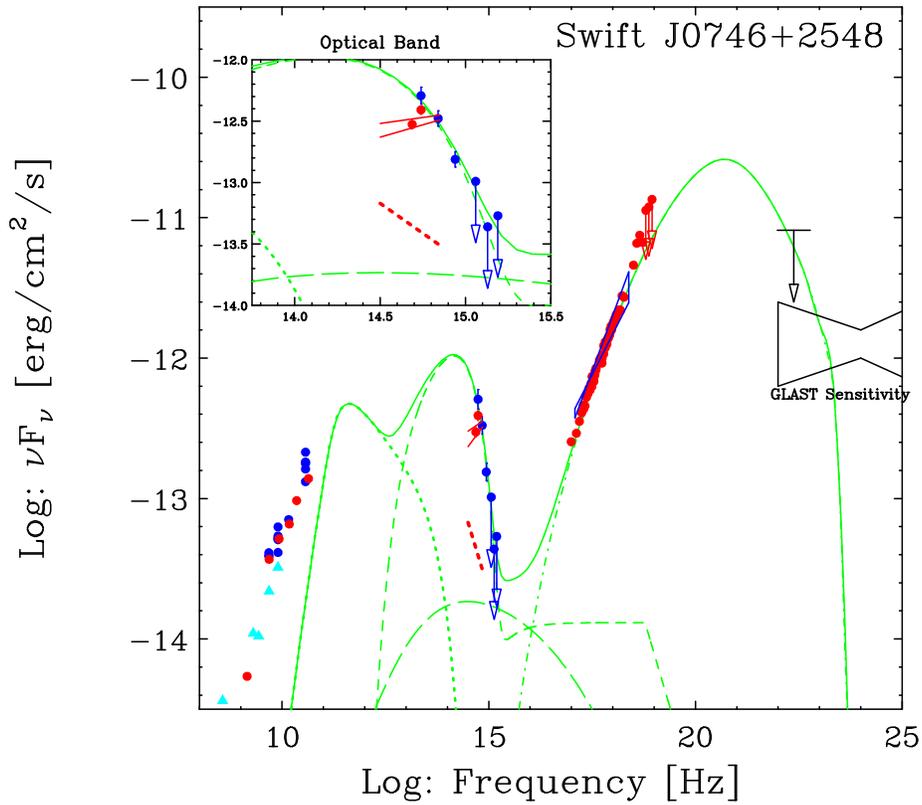}
\caption{Overall SED of J0746 constructed with multiband data obtained 
during the November 2005 campaign (red and blue).
The red filled circles and lines show the observation results presented in this paper:
radio (VLA), optical-UV(NOT; filled circles and HET; lines) and X-ray ($Suzaku$). 
For the HET data, the difference spectrum between the two epochs is shown in the red dotted line.
The blue filled circles and lines show the data presented by \citet{sam06}: 
radio (Metsahovi and UMRAO), optical-UV ($Swift$ UVOT) and X-ray ($Swift$ XRT). 
The data plotted with cyan triangles are from NED, 
while the GeV upper limit shows the EGRET data analyzed by \citet{sam06}.  
The green solid line shows the jet continuum calculated 
with the jet emission model described in $\S$ 4.3, as a sum of various emission components: 
synchrotron (dotted line), blue bump (dashed line), SSC (long dashed line) and ERC (dot dashed line).
Moreover, the sensitivity of one year GLAST observation is also plotted for reference.}
  \label{fig:nufnu}
\end{center}
\end{figure}

Figure~\ref{fig:nufnu} shows the spectral energy distribution (SED) of
J0746 during the 2005 November campaign. 
Blue and red represent simultaneous data of the radio, UV, optical and X-ray 
observations. 
Historical data taken from radio (NED), 
and $\gamma$-ray \citep[EGRET upper limit;][]{sam06} observations are also plotted 
in cyan and black, respectively. 
Figure~\ref{fig:nufnu} implies that the synchrotron component of 
J0746 most likely peaks around $\sim 10^{11}-10^{12}$ Hz in the observer frame. 
Meanwhile, $Swift$ UVOT data show the steep optical-UV emission 
as the high-energy tail of the ``blue bump'' 
which is thought to be produced via thermal emission by the accretion disk
and/or corona near the central black hole of J0746 \citep{sun89}. 
Apparently, these optical-UV data do not join smoothly with 
the X-ray-to-$\gamma$-ray spectrum, which is likely due to the nonthermal 
External Compton jet radiation (ERC) or due to synchrotron-self-Compton 
emission (SSC) \citep[e.g.,][]{bal02}.

In order to model the SED of J0746, we applied the synchrotron-inverse Compton (IC) emission 
model described in \citet{mar03},
where both synchrotron 
and external photons are considered as seed radiation fields contributing 
to the IC process. 
The electron distribution is modeled as a smoothed broken power-law; 
\begin{equation}
n_e(\gamma) = K \gamma^{-n_1} \left(1 + \frac{\gamma}{\gamma_{\rm brk}} \right)^{n_1 - n_2},
\end{equation}
where $K$ (cm$^{-3}$) is a normalization factor, $n_1$ and $n_2$ are the spectral 
indices below and above the break Lorentz factor $\gamma_{\rm brk}$.
The electron distribution extends within the limits 
$\gamma_{\rm min} < \gamma < \gamma_{\rm max}$. 

We assume that the blazar radiation originates in a region located at a distance $r$ from the 
black hole, well within the Broad Line Region but sufficiently far above the accretion disk
that the radiation energy density from the latter can be neglected.
The external radiation field can then be simply modeled,
\begin{equation}
U_{\rm diff} \simeq \frac{L_{\rm BLR}}{4 \pi r_{\rm BLR}^2 c},
\end{equation}
where $r_{\rm BLR}$ is the size of the broad-line-region.

Figure~\ref{fig:nufnu} shows the best-fit model for J0746 data, 
combining of the synchrotron, the SSC, and the ERC components \citep{tav08}, 
plus the blue bump emission. 
The spectrum can be completely fitted with the model parameters; 
the emission region is modeled as a sphere with radius $R = 3.2\times10^{16}$ 
cm moving with a bulk Lorentz factor $\Gamma = 20$ at an angle $\theta = 0.05$ rad 
between the line of sight and the jet axis, and filled by tangled magnetic field $B=1.8$ G 
and relativistic electrons.
The Doppler beaming factor is
$\delta \equiv 1/[\Gamma (1 - \beta \cos{\theta})] \sim 20$. 
Parameters of the electron distribution are 
$n_1=1.34$, $n_2=3.8$, $\gamma_{\rm min} = 1$, 
$\gamma_{\rm brk} = 27$, $\gamma_{\rm max} = 10^3$, respectively. 
The size of the BLR is assumed to be $r_{\rm BLR}=6.5\times10^{17}$ cm, and the 
luminosity of broad emission lines $L_{\rm BLR}=2\times10^{45}$~ergs~s$^{-1}$. 
The disk blue bump has a luminosity $L_{\rm disk} \simeq 1.8\times10^{47}$~ergs~s$^{-1}$ 
and temperature $kT_{\rm UV} = 10$~eV  (redshifted temperature 2.5~eV).  

\section{Discussion}

\subsection{$Suzaku$ results of J0746}

In previous sections, we presented temporal and spectral analysis of $Suzaku$ observation 
of J0746 in 2005 November. 
Using the high-sensitivity, broadband instruments onboad $Suzaku$, we found 
the following characteristics of J0746: 
(1) The variability amplitude of soft (0.5--2~keV) and hard (2--10~keV) bands are 
both $F_{\rm var,soft} \sim F_{\rm var,hard} \simeq 0.03$. There seems to be 
no significant energy-dependence of the variability. 
(2) The observed X-ray spectrum is well-described by 
a hard power-law ($\Gamma_{\rm ph} = 1.17$) 
with the Galactic absorption. 
Thanks to the good photon statistics and spectral response of $Suzaku$ XIS, we clearly 
confirmed that J0746 has an intrinsically hard spectrum and can exclude the possibility
that the spectral hardening results from the excess absorption as reported 
by \citet{sam06}.
Such differences of the spectrum between $Swift$ and $Suzaku$ are probably due to 
(1) low statistics of $Swift$ XRT compared to $Suzaku$, and 
(2) \citet{sam06} combining the X-ray spectra obtained at 
4 different epochs, with different exposures.
The observed photon index is extremely hard, similar to those observed 
in several high-luminosity blazars \citep[e.g.,][]{tav00}.
As long as the X-ray emission is due to the low-energy end of the ERC spectral component, 
the observed photon index $\Gamma_{\rm ph} = 1.17$ corresponds to the electron distribution 
$n_e(\gamma) \propto\gamma^{-1.34}$, where $\gamma$ 
is the Lorentz factor of the ultrarelativisitic electrons. 
A likely explanation of such a flat electron distribution 
is discussed by \citet{sik02} who assume a two-step acceleration process: 
the harder portion is produced by a pre-acceleration mechanism,  
e.g., involving instabilities driven by shock-reflected ions \citep{hos92} or 
magnetic reconnection \citep{rom92}, 
while the high energy tail by the standard first-order Fermi process operating over the shock front.

\subsection{Constraint on Bulk Compton emission}

As the cold electrons/positrons, before reaching the blazar dissipative site,
are transported from the black hole vicinity by a jet with a bulk Lorentz
factor $\Gamma_{\rm jet} \sim$~10--20,
they upscatter the accretion disk and broad emission line photons to energies
\begin{equation}
h \nu_{\rm BC,obs} \simeq \Gamma_{\rm jet} \delta_{\rm jet} h\nu_{\rm diff}/(1+z) \, 
\end{equation}
where $h\nu_{\rm diff} \sim 10$~eV. This is expected to produce a hump in the
X-ray spectra of blazars with luminosity
\begin{equation} \label{eq:lbc}
L_{BC} \simeq N_e \vert dE_e/dt \vert (\delta_{\rm jet}^3/\Gamma_{\rm jet})  \simeq \frac{4}{3} \sigma_T U_{BLR} r_{BLR} \dot N_e  \Gamma_{\rm jet} \delta_{\rm jet}^3 ,
\end{equation}
where $\vert dE_e/dt \vert = (4/3) c \sigma_T U_{BLR} \Gamma_{\rm jet}^2$ and 
$N_e \simeq \dot N_e r_{BLR}/c$ is the number of electrons
enclosed in the jet within a distance range corresponding with the scale of
the broad emission line region.

\begin{figure}
\begin{center}
\epsscale{0.80}
\plotone{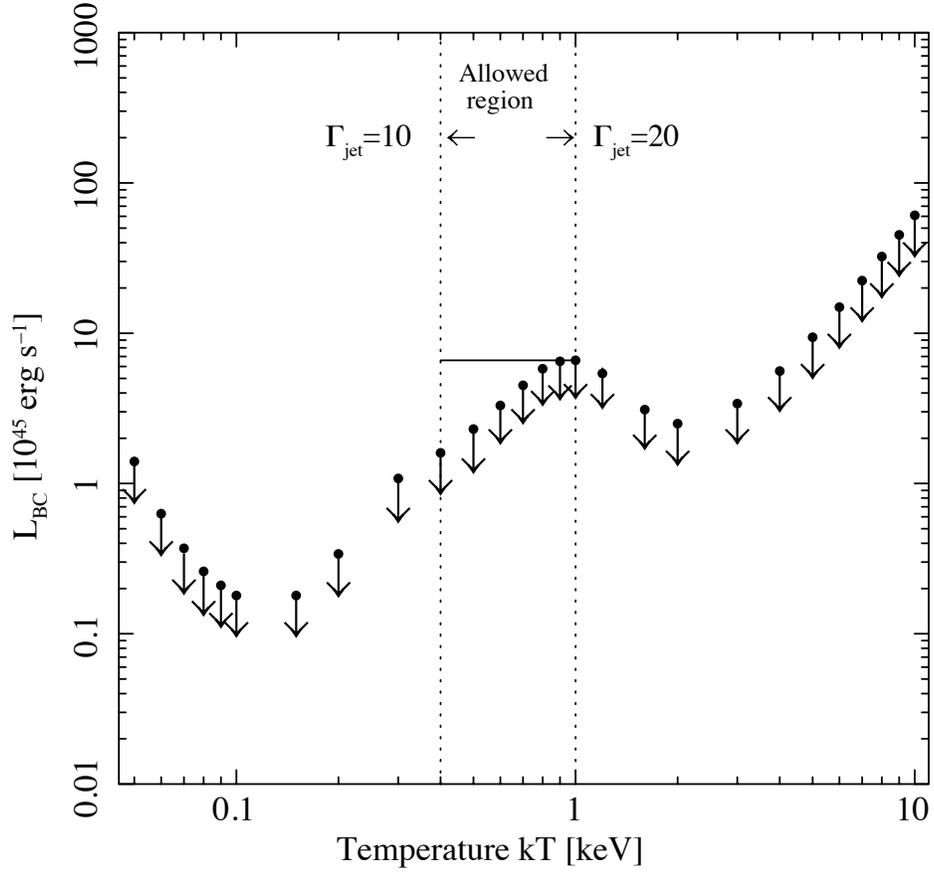}
\caption{The 1 $\sigma$ upper limit of the BC component luminosity estimated 
from $Suzaku$ data fitting with power-law + black-body model.
  \label{fig:LBC}}
\end{center}
\end{figure}

\begin{figure}
\begin{center}
\epsscale{0.80}
\plotone{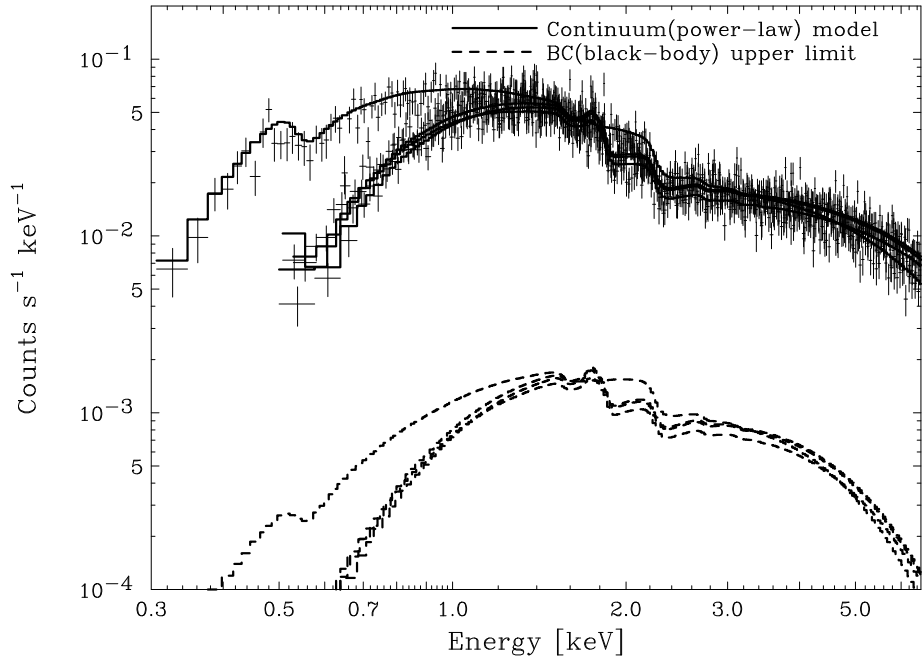}
\caption{A comparison between the model (continuum + bulk Compton component) and the data. 
As the the bulk-Compton component, a black-body with a temperature of 1.0~keV is assumed, and 
the 1 $\sigma$ upper limit is plotted.} 
 \label{fig:comparision}
\end{center}
\end{figure}

For our observation of J0746, the soft X-ray excess which would indicate the 
BC feature is not detected.
However, since the $Suzaku$ observation of J0746 was performed in a relatively low state
with an average flux of $F_{\rm 2-10 keV} \sim 3\times 10^{-12}$ {\rm erg cm}$^{-2}$ {\rm s}$^{-1}$,
we can put a stringent upper limit on the BC emission.
The limit is presented in Figure~\ref{fig:LBC}. It is obtained 
using $Suzaku$ data fitted with power-law determined in \S~4.2 plus black-body
approximation of the bulk-Compton component. 
As an example, a comparison between the model and the data is shown in Figure~\ref{fig:comparision}, and,
some of the fitting results are also listed in Table~\ref{tab:spec} (model 3-1 and model 3-2). 
In Figure~\ref{fig:LBC}, the marked 'allowed region'
corresponds with a temperature range $kT$~=~0.40--1.0~keV which, in turn,
corresponds with $\Gamma_{\rm jet} = \delta_{\rm jet}$~$\sim$~10--20. In this region the upper 
limit of BC luminosity is  $L_{\rm BC} \leq 6.6\times10^{45}$~erg~s$^{-1}$ .

It should be noted here that in the case of the popular internal shock model
\citep[e.g.,][]{spa01} the bulk-Compton radiation is produced by 
two cold inhomogeneities/shells.
In this case production of any nonthermal flare by the internal shock 
should be proceded  by a pair of X-ray precursors: one produced by 
a faster shell at larger energies and lasting shorter; and one produced 
by a slower shell at lower energies and lasting longer.
Radiative bulk-Compton features  from such systems are very complex, 
are variable and depend on the model details \citep{mod04}. However, 
a small amplitude of variability in J0746 (see Figure~\ref{fig:xis_lc}) suggests that if 
the primary dissipative events are driven by internal shocks, what we observe 
is an overlap of radiation contributed by several shocks. Then the upper 
limits for the bulk-Compton emission
calculated using the 'steady state' may be a reasonable approximation.

\subsection{Constraint on particle content in the jet of J0746}
Noting that the energy flux carried by the cold electrons is
\begin{equation} \label{eq:le}
L_{e, \rm cold} \simeq n_e m_e c^3 \Gamma^2_{\rm jet} \pi R_{\rm jet}^2 \equiv \dot N_e m_e c^2 \Gamma_{\rm jet},
\end{equation}
where $R_{\rm jet}$ is the cross-section radius of a jet, one can find after combining Eq.~(\ref{eq:lbc}) and Eq.~(\ref{eq:le}) that
\begin{equation}
L_{\rm BC} \simeq \frac{4 \sigma_{\rm T}}{3 m_e c^2} ~U_{\rm BLR} ~r_{\rm BLR} ~\delta_{\rm jet}^3 ~L_{e, \rm cold}.
\end{equation}
For the upper limit for $L_{\rm BC}$ given by $Suzaku$ data (see \S~{5.2}) this gives
\begin{equation}
L_{e,\rm cold}  \leq 1.0\times10^{44} 
~\biggl(\frac{r_{\rm BLR}}{6.5\times10^{17}~{\rm cm}} \biggr) 
~\biggl(\frac{\delta_{\rm jet}}{20} \biggr)^{-3} 
~\biggl(\frac{L_{\rm BLR}}{1.8\times10^{45}~{\rm erg}~{\rm s}^{-1}} \biggr)^{-1} 
~{\rm erg} ~{\rm s^{-1}}.
\end{equation}

Meanwhile, the ERC modeling of J0746 presented in the previous section ($\S$~4.3) 
implies the jet power carried by the ultrarelativisitic (non-thermal, or 'hot') electrons 
$L_{e,\rm hot} \simeq \dot N_e m_ec^2 \bar \gamma_e \Gamma_{\rm jet} \sim 4 \times 10^{45}$~erg~s$^{-1}$,
where $\bar \gamma_e$ is the average random Lorentz factor of electrons/positrons.
However, if a jet is free of protons and the only source of the energy is 
the bulk energy of cold pairs, then from the energy conservation 
one can deduce that $L_{e, \rm hot} < L_{e, \rm cold}$. This is in a clear disagreement 
with the obtained upper limit for $L_{e,\rm cold}$ and the model value of $L_{e,\rm hot}$. 

Such situation may be avoided if one assumes that there are cold protons 
which carry significant power $L_p > L_{e, \rm hot} \gg L_{e,\rm cold}$. In this case 
$L_{e,\rm hot}/(L_{e,\rm cold}+L_{p,\rm cold}) \lesssim 1$, and, provided that jet kinetic luminosity 
$L_{\rm jet} \simeq L_{p,\rm cold}$, the pair content reads as 
\begin{equation}
\frac{n_e}{n_p} 
= \frac{m_p}{m_e} \frac{L_{e,\rm cold}}{L_{p, \rm cold}}
\simeq \frac{m_p}{m_e} \frac{L_{e,\rm cold}}{L_{\rm jet}}.
\end{equation}
Noting that the luminosity of the observed high energy ($\gamma$-ray) emission can be related 
to the jet kinetic luminosity via the relation 
$L_{\gamma} \simeq \eta_{\gamma} (\delta_{\rm jet}^3 / \Gamma_{\rm jet}) L_{\rm jet}$, 
where $\eta_{\gamma}$ is the efficiency of the high energy $\gamma$-ray production, we finally find 
that the upper limit for the pair content of the J0746 jet is
\begin{equation}
\frac{n_e}{n_p} \leq 7.3 \times 
~\biggl(\frac{\eta_{\rm \gamma}}{0.1}\biggr) 
~\biggl(\frac{r_{\rm BLR}}{6.5\times10^{17}~{\rm cm}}\biggr)
~\biggl(\frac{\Gamma_{\rm jet}}{20} \biggr)^{-1} 
~\biggl(\frac{L_{\rm BLR}}{1.8\times10^{45}~{\rm erg}~{\rm s}^{-1}} \biggr)^{-1} 
~\biggl(\frac{L_{\rm \gamma}}{10^{48}~{\rm erg}~{\rm s}^{-1}} \biggr)^{-1}.
\end{equation}
For J0746, we only have an upper limit on the gamma-ray flux of 
$L_{\gamma} \sim 10^{48}$~erg~s$^{-1}$.  With this, and $L_{\rm BLR} \sim 1.8\times10^{45}$~erg~s$^{-1}$, 
we obtained $n_e \leq 7.3 ~(\eta_{\gamma}/0.1) ~n_p$. 
This may indicate a rather low pair content in quasar jets. However, it 
should be noted that J0746 has an exceptionally hard X-ray spectrum. 
For  blazars with softer X-ray spectra a lack of bulk-Compton features
put weaker constraints, $n_e/n_p \le $ tens, But the inertia of such jets 
is still dominated by protons \citep{sik00}.

\section{Summary}

We have presented a detailed analysis of $Suzaku$ observation for the radio-loud 
quasar J0746 at $z=2.979$ in 2005 November. Our results are the following:

\begin{enumerate}
\item The variability amplitude of soft and hard bands as measured by $Suzaku$ 
is equivalent and there is no significant energy-dependence of the variability, 
in contrast to the much larger (factor of 2) variability reported from the Swift data.
\item The observed X-ray spectrum of J0746 is well-described by a single, extremely hard 
power-law ($\Gamma_{\rm ph} = 1.17$) with the Galactic absorption;  
we do not require spectral hardening at the lowest energies seen by \citet{sam06}.  
With this, we can exclude excess absorption (which would otherwise have to be
rapidly variable!) to cause 
the spectral hardening, one of the possibilities considered by \citet{sam06}.
\item A lack of bulk Compton features in the X-ray spectra indicates 
a low electron-positron pair content and strong dominance jet inertia by protons.
\end{enumerate}

\acknowledgments
We thank the anonymous referee for her/his valuable comments that helped
to improve this paper.

The National Radio Astronomy Observatory (NRAO) is a facility of the
National Science Foundation operated under cooperative agreement by
Associated Universities, Inc. NRAO is thanked for the provision of
Target of Opportunity time for the observations, and Barry Clark is
thanked for assistance is the preparation of observing files.  
The research described here we supported in part by the 
Department of Energy contract to SLAC no. DE-AC3-76SF00515, and NASA 
grant to Stanford University no. NNX07AB05G.  
Based on observations made with the Nordic Optical Telescope, operated
on the island of La Palma jointly by Denmark, Finland, Iceland,
Norway, and Sweden, in the Spanish Observatorio del Roque de los
Muchachos of the Instituto de Astrofisica de Canarias.

\clearpage

\clearpage

\begin{deluxetable}{cccc}
\tablecaption{2005 Suzaku observation log of J0746. \label{tab:obs}}
\tablewidth{0pt}
\tablehead{
\colhead{Start (UT)} & \colhead{Stop (UT)} & \colhead{Exposure (ks)} & \colhead{Exposure (ks)} \\ 
\colhead{}           & \colhead{}          & \colhead{XIS} & \colhead{HXD/PIN}
}
\startdata
Nov. 04 08:20 2005 & Nov. 06 14:04 2005 & 100.5 & 74.0 \\
\enddata
\end{deluxetable}

\begin{table}
  \begin{center}
  \caption{NOT photometric observations of J0746.\label{tab:notobs}}
    \begin{tabular}{ccccccc}
\tableline\tableline
Band & Flux (mag) & std\tablenotemark{a} & photerr\tablenotemark{b} & calibration\tablenotemark{c} & exposure (second) & time (UT) \\ 
\tableline
R-band & 18.888 & 0.006 & 0.011 & (1) & 200 & 2005 November 5  \\
       & 18.937 & 0.017 & 0.011 & (2) &     & 05:38:37 \\
       \tableline
V-band & 19.232 & 0.024 & 0.010 & (1) & 200 & 2005 November 5 \\
       & 19.233 & 0.005 & 0.010 & (2) &     & 05:44:05  \\
\tableline
    \end{tabular}
    \\
    \tablenotetext{a}{Standard deviation of the target brightness estimates.}
    \tablenotetext{b}{The apphot error estimate for the target.}
    \tablenotetext{c}{ 
    (1): using the published SDSS g'r' magnitudes and transforming these magnitudes to V and R.
    (2): using Landolt standard stars (PG2213$-$006 and Mark\_A).}
  \end{center}
\end{table}

\begin{table}
\begin{center}
\caption{VLA observations of J0746.\label{tab:vlaobs}} 

\begin{tabular}{cc}
\tableline\tableline
Frequency (GHz) & Flux density (Jy) \\ 
\tableline
 1.425    &           0.38 \\
 4.860    &           0.76 \\
 8.460    &           0.61 \\
14.940    &           0.44 \\
22.460    &           0.43 \\
43.340    &           0.32 \\
\tableline
    \end{tabular}
  \end{center}
\end{table}

\clearpage

\begin{landscape}
\begin{deluxetable}{llccccc}
\tabletypesize{\scriptsize}
\tablecaption{Results of the spectral fits to the $Suzaku$ spectra.\label{tab:spec}}
\tablewidth{0pt}
\tablehead{
\colhead{Component} & \colhead{Parameter} & \colhead{model 1} & \colhead{model 2} & \colhead{model $2^{\prime}$} &
\colhead{model 3-1} & \colhead{model 3-2} 
}
\startdata
Absorption & $N_{\rm H}$ ($10^{20}$cm$^{-2}$)                       & 4.04 (fixed)  & 4.04 (fixed)  & 4.89$\pm$0.50 & 4.04 (fixed)      & 4.04 (fixed)      \\
Power-law  & $\Gamma_{\rm ph}$                                      & 1.17$\pm$0.01 & 1.18$\pm$0.01 & 1.20$\pm$0.01 & 1.17 (fixed)      & 1.17 (fixed)      \\
           & $F_{2-10~\rm keV}$ ($10^{-12}$~erg~s$^{-1}$~cm$^{-2}$) & 3.10$\pm$0.02 & 3.07$\pm$0.03 & 3.06$\pm$0.04 & 3.08$\pm$0.02     & 3.05$\pm$0.04     \\
Constant   & XIS 0                                                  & 1.00 (fixed)  & 1.00 (fixed)  & 1.00 (fixed)  & 1.00 (fixed)      & 1.00 (fixed)      \\
           & XIS 1                                                  & 1.00 (fixed)  & 0.91$\pm$0.01 & 0.91$\pm$0.01 & 1.00 (fixed)      & 1.00 (fixed)      \\
           & XIS 2                                                  & 1.00 (fixed)  & 1.04$\pm$0.01 & 1.04$\pm$0.01 & 1.00 (fixed)      & 1.00 (fixed)      \\
           & XIS 3                                                  & 1.00 (fixed)  & 1.05$\pm$0.01 & 1.05$\pm$0.01 & 1.00 (fixed)      & 1.00 (fixed)      \\
           & HXD/PIN                                                & 1.15 (fixed)  & 1.15 (fixed)  &  1.15 (fixed) & 1.15 (fixed)      & 1.15 (fixed)      \\
Black-Body & Temperature (keV)                                      & -             & -             & -             & 0.4 (fixed)       & 1.0 (fixed)       \\
 (Bulk-Compton) & Luminosity($10^{45}$~erg~s$^{-1}$)                & -             & -             & -             & 0.8 (0.0--1.6)    & 3.9 (1.0--6.6)    \\
 \tableline
$\chi^{2}$/d.o.f ($\chi_{\rm red}^2$) &                             & 1238/1112 (1.11) & 1113/1109 (1.00) & 1110/1108 (1.00) & 1238/1112 (1.11) & 1237/1112 (1.11) \\
\enddata
\tablecomments{Errors correspond to 1~$\sigma$ confidence level.}
\end{deluxetable}
\clearpage
\end{landscape}

\end{document}